\documentclass[sigconf]{acmart}

\usepackage{graphicx}
\usepackage{textcomp}
\usepackage{xcolor}
\usepackage{xspace}
\usepackage[acronym,nohypertypes={acronym}]{glossaries}
\usepackage{enumitem}
\usepackage{orcidlink}
\usepackage{algorithm}
\usepackage{algpseudocode}
\algrenewcommand\algorithmicrequire{\textbf{Input:}}
\algrenewcommand\algorithmicensure{\textbf{Output:}}

\usepackage{amsmath}
\usepackage[textwidth=17mm]{todonotes}

\newcommand{\customtodo}[4]{
    \todo[color=#2,inline,size=\small]{
        \ifx&#3&
        \textbf{#1} #4
        \else
        \textbf{#1$\Rightarrow$#3} #4
        \fi
    }
}

\newcommand{\sys}{\textit{Time-Shared Computing}\xspace}

\newacronym{ict}{\textsc{ICT}}{\textit{Information and Communication Technologies}}
\newacronym{iaas}{\textsc{IaaS}}{\textit{Infrastructure-as-a-Service}}
\newacronym{mlaas}{\textsc{MLaaS}}{\textit{ML-as-a-Service}}
\newacronym{faas}{FaaS}{\textit{Function as a Service}}
\newacronym{pm}{PM}{\textit{Physical Machine}}
\newacronym{ai}{AI}{\textit{Artificial Intelligence}}
\newacronym{hpc}{HPC}{\textit{High-Performance Computing}}
\newacronym{vm}{VM}{\textit{Virtual Machine}}
\newacronym{cpu}{CPU}{\textit{Central Processing Unit}}
\newacronym{gpu}{GPU}{\textit{Graphics Processing Unit}}
\newacronym{dc}{DC}{\textit{Data Center}}
\newacronym{pue}{PUE}{\textit{Power Usage Efficiency}}
\newacronym{vcpu}{vCPU}{\textit{virtual CPU}}
\newacronym{vram}{vRAM}{\textit{virtual RAM}}
\newacronym{slo}{SLO}{\textit{Service-Level Objective}}
\newacronym{sla}{SLA}{\textit{Service-Level Agreement}}
\newacronym{itmt}{ITMT}{\textit{Intel Turbo Boost Max Technology}}
\newacronym{cfs}{CFS}{\textit{Completely Fair Scheduler}}
\newacronym{eevdf}{EEVDF}{\textit{Earliest Eligible Virtual Deadline First}}
\newacronym{rss}{RSS}{\textit{Resident Set Size}}
\newacronym{wss}{WSS}{\textit{Working Set Size}}
\newacronym{smt}{SMT}{\textit{Simultaneous Multithreading}}
\newacronym{mpc}{M/C}{\textit{Memory per Core}}
\newacronym{numa}{NUMA}{\textit{Non-Uniform Memory Access}}
\newacronym{qos}{QoS}{\textit{Quality of Service}}
\newacronym{pdu}{PDU}{\textit{Power Distribution Unit}}
\newacronym{bmc}{BMC}{\textit{Baseboard Management Controller}}
\newacronym{rapl}{RAPL}{\textit{Running Average Power Limit}}
\newacronym{dsb}{DSB}{DeathStarBench}
\newacronym{dbaas}{DBaaS}{Database as a Service}
\newacronym{dbms}{DBSM}{Database Management System}
\newacronym{wue}{WUE}{Water Usage Effectiveness}
\newacronym{vfio}{VFIO}{Virtual Function I/O}
\newacronym{iommu}{IOMMU}{Input-Output Memory Management Unit}
\newacronym{mig}{MIG}{\textit{Multi-Instance GPU}}
\newacronym{mps}{MPS}{\textit{Multi-Process Service}}
\newacronym{gi}{GI}{\textit{GPU instance}}
\newacronym{ci}{CI}{\textit{Compute Instance}}
\newacronym{sm}{SM}{\textit{Streaming Multiprocessor}}
\newacronym{ipmi}{IPMI}{\textit{Intelligent Platform Management Interface}}
\newacronym{tdp}{TDP}{\textit{Thermal Design Power}}
\newacronym{vps}{VPS}{\textit{Virtual Private Server}}
\newacronym{pci}{PCI}{\textit{Public Cloud Infrastructure}}
\newlist{questions}{enumerate}{2}
\setlist[questions,1]{label=RQ\arabic*.,ref=RQ\arabic*}
\setlist[questions,2]{label=(\alph*),ref=\thequestionsi(\alph*)}
\AtBeginDocument{%
  }

\setcopyright{none}
\acmDOI{10.48550/arXiv.2507.19287}
\acmISBN{}

\acmConference[LIMITS '25]{11th Workshop on Computing within Limits}{June 26--27, 2025}{Online}

\begin{document}

\title{The Case for Time-Shared Computing Resources}
\author{Pierre Jacquet}
\orcid{0009-0002-7988-8550}
\affiliation{%
  \institution{École de Technologie Supérieure, Université du Québec}
  \city{Montréal}
  \country{Canada}
}
\email{pierre.jacquet@etsmtl.ca}

\author{Adrien Luxey-Bitri}
\orcid{0000-0003-1777-307X}
\affiliation{%
  \institution{Université de Lille, Inria, CNRS, UMR 9189 CRIStAL}
  \city{Villeneuve d’Ascq}
  \country{France}
}
\affiliation{%
  \institution{Deuxfleurs Association}
  \country{France}
}
\email{adrien.luxey@inria.fr}
\renewcommand{\shortauthors}{Jacquet et al.}

\begin{abstract}

The environmental impact of \gls{ict} continues to grow, driven notably by increasing usage, rebound effects, and emerging demands. 
However, despite the virtual nature of its services, the sector remains inherently constrained by its materiality and cannot rely on an infinite pool of resources.
As a result, the wide variety of supported services may need to be managed under stricter limits within hosting facilities in the future.

Contrary to common assumptions, we show that tenants typically do not share computing resources—even in environments commonly perceived as mutualized, such as cloud platforms.
Time-sharing has been progressively phased out for reasons of performance, security, predictability, and, perhaps more importantly, due to the decreasing cost of computing resources. 

This paper advocates for managing fewer physical resources by improving resource sharing between tenants.
It represents a paradigm shift, moving beyond traditional time-sharing at the hardware level to a higher abstraction. 
This approach entails "doing with fewer resources" under conditions of "reduced performance".
Nonetheless, enhancing the mutualization of infrastructure can reduce cluster sizes (through consolidation) and improve energy efficiency, with gains related to the accepted performance trade-off, a situation potentially more socially acceptable than eliminating services.
We review the current state of the art, identify challenges and opportunities, propose interpretations of \sys, and outline key research directions.
\end{abstract}

\glsresetall



\maketitle
\glsresetall

\section{Introduction}\label{sec:intro}

Different paradigm shifts have occurred throughout the history of \gls{ict}, driven by technological innovation and changes in resource costs.
Distributed systems progressively replaced initial centralized and constrained architectures to improve scalability and fault tolerance.
As a result, computing resources became increasingly dedicated to specific applications, moving away from earlier, more constrained settings where resource sharing was a necessity.
The authors of this paper argue that today's architectures may once again encounter physical limits, as \gls{ict} faces growing constraints on resources.

Different types of constraints can be identified, the most visible being the electricity availability, identified by the pressure that \gls{ict} places on existing power grids.
Since 2017, electricity consumption from data centres has increased by approximately 12\% per year—more than four times faster than the overall global electricity demand~\cite{ieaia}.
Such an exponential trajectory raises concerns about the ability of existing infrastructures to keep pace with the expansion of digital services.

Another physical constraint stems from the availability of critical minerals, which are essential to the production and operation of \gls{ict} infrastructures~\cite{Bessai2023Fit,cerf:hal-04709741}. 
The competition for these resources is increasingly intense as they are also crucial to other strategic sectors.

From a climate perspective, \gls{ict} was responsible for approximately 2 to 3\% of global greenhouse gas emissions~\cite{FREITAG2021100340}—a figure that may increase due to the growing demand for data, cloud services, and digital infrastructure.
Additional constraints may arise from carbon policies aimed at curbing this trend.

In response to these limitations, fewer hardware resources may be available in the future.
Rather than simply enduring these limits, they should be anticipated and actively managed.
Yet, the future in which digital infrastructures must operate under real, material limits will differ significantly from the early mainframe era, not only due to the knowledge accumulated and the greater capabilities of modern infrastructure, but also because of our society's deep reliance on these digital systems~\cite{bugeau2023digital}.

Part of the solution may lie in more efficient sharing of existing physical resources, that is, by using systems capable of supporting multiple tenants on the same computing resources (i.e., sharing their available time between applications).
While pooling end-user terminals presents practical challenges, and networks are already de facto shared, we argue that the most promising opportunities reside in the mutualization of server-side resources.

In hosting facilities, sharing is mostly limited to spatial partitioning (computing resources are virtually divided into smaller subsets through virtualization), while time-sharing is often overlooked.
Although it is typically addressed as a system-level problem (e.g., how processes are scheduled on physical components), we argue that it should also be considered from a higher perspective: How can tenant applications consume the minimal amount of resources, notably by sharing them with others?
Since \gls{ict} has never had to operate with fewer transistors~\cite{rupp2021microprocessor}, this paper explores what \sys could be under such limits.


We first review the current state of hosting mutualization in Section~\ref{sec:rw}.
In Section~\ref{sec:principle}, we outline the core principles of \sys{}.
We then identify key challenges the community must address across different scenarios, before concluding the paper in Section~\ref{sec:conclusion}.

\section{State of the Art}\label{sec:rw}

In this section, we briefly review how resource sharing has historically been implemented in computing infrastructures and how it is currently approached in modern environments.

\subsection{From constrained environments...}

The concept of \sys is almost as old as software itself. 
The transition to programmable systems allowed a single physical machine to be used for different purposes without modifying its hardware architecture.

One of the earliest examples of shared computing can be found in mainframes. 
Due to their high cost and centralized nature, mainframes were shared within large organizations, implementing what is arguably the earliest form of sharing: sequential batch processing, where jobs were executed one after another (e.g., processing punch cards in sequence).
This may be seen as a first naive time-shared principle.

In the 1960s, job switching was introduced to improve efficiency, particularly by utilizing idle CPU cycles while waiting for slow I/O operations~\cite{timeshared}.

This led to the development of modern time-sharing mechanisms, where jobs are regularly switched to create the illusion of concurrency. 
This was a revolutionary step, as it enabled multiple users to interact with a single machine simultaneously.
Time-sharing techniques eventually evolved into virtualization, a concept that allowed a single physical machine to host multiple virtual ones, improving isolation between users and optimizing hardware utilization~\cite{ibmVM,5388296}.

The introduction of multi-core \gls{cpu} architectures marked another milestone (such as with IBM System/360 Model 65MP in 1965), as it enabled different jobs to run concurrently on separate computing resources.

A natural evolution from this was the emergence of distributed architectures, where multiple interconnected servers worked together (e.g., gossip protocol~\cite{demers1987epidemic}). 
This shift was enabled by advancements in networking and the decreasing cost of \gls{ict} equipment~\cite{priceIT}.

Initially, distributed computing followed a client-server model, where dedicated servers handled specific tasks. 
This approach moved away from traditional multi-tenant resource sharing (where different users shared the same system) toward a single-tenant perspective, where resources were shared among processes belonging to the same organization.

In parallel, research on cluster computing and grid computing explored how groups of machines could collaborate to solve large-scale problems. 
These paradigms primarily targeted \gls{hpc} applications rather than general-purpose computing.

For many years, large enterprises operated their physical servers (sometimes rented) but the rising complexity of infrastructure management led to the adoption of virtualized and shared hosting environments, giving birth to cloud computing.

\subsection{...To modern ICT}

Cloud computing introduced large-scale mutualization, where \glspl{dc} resources are shared by multiple tenants.
Unlike colocated \glspl{dc}, which just shares infrastructure elements such as the building, the cooling, and power supply between bare-metal servers, cloud computing also shares servers (and part of their resources) between clients.
Virtualization allowed multiple users to share the same physical servers, improving efficiency and flexibility.

However, cloud computing did not fully restore the practice of sharing computing resources. 
Today, \glspl{vcpu} are typically provisioned in a 1:1 mapping with physical CPU cores in cloud platforms~\cite{awsOC,258979}, limiting platform usage~\cite{10.1145/3132747.3132772}. 

While oversubscription is documented in \gls{dc} production environments~\cite{215941}, even from actors present in cloud infrastructures~\cite{tirmazi2020borg,10.1145/3447786.3456259}, it is not common for workloads other than internal.
Different reasons may be identified, notably performance predictability, \glspl{sla}, security, and reputation.
On the security front, co-locating tenants on shared physical resources increases the risk of side-channel attacks, particularly those exploiting CPU cache mechanisms~\cite{maurice:hal-04679284}.

As a result, \textbf{spatial}-sharing is the norm, as a server is divided through virtual instances.
But \textbf{time}-sharing is the exception confined to low-cost cloud offerings~\cite{ovhDiscovery,burstazure,burstaws}.
We argue that only time-sharing can truly mutualize all resources, as spatial sharing only shares non-reserved resources: multiple \glspl{vm} may share the same motherboard but will use different computing cores, memory pages, and disk blocks.\\

In summary, the practice of time-sharing computing resources across different jobs has vastly diminished since the rise of distributed architectures.
Today, it is primarily left to the kernel to decide how processes are allocated to cores rather than being treated as a multi-tenant problem at a higher level.

This paper advocates for a shift in perspective: computing resources should not only be shared for cost efficiency but also as a deliberate strategy to reduce the environmental impact of IT infrastructure.

\section{Principle}\label{sec:principle}

\begin{table*}[hbt!]
    \caption{Classification of Time-Shared Policies for Applications}\label{tab:classification}
    \centering
    \begin{tabular}{|c|c|c|c|}
        \hline
        \bf Type & \bf Service & \bf Inconvenience \\ 
        \hline
        Sequential Time-Sharing & Virtual resource & Delay deployment or vertical elasticity\\
        \hline
        Concurrent Time-Sharing & Virtual resource & Potential performance reduction\\
        \hline
        Sequential Time-Sharing & Functional unit & Delay execution\\
        \hline
        Concurrent Time-Sharing & Functional unit & Potential performance reduction\\
       \hline
    \end{tabular}
\end{table*}

As discussed earlier, \sys is a fundamental concept in software, traditionally managed by the operating system, which schedules processes on available computing resources. 
However, resource sharing is not a primary design objective for most developers outside of kernel and systems programming. 

This paper advocates for \sys to become a core objective of modern applications hosting.
There are several compelling reasons to adopt this perspective:

\begin{itemize}
\item \textbf{Reducing infrastructure expansion:} Sharing computing resources minimizes material needs.
\item \textbf{Improving energy efficiency:} Maximizing the utilization of existing platforms leads to higher energy efficiency~\cite{jacquet2025cinergy}, as underutilized resources still consume power even when idle.
\item \textbf{Enhancing expertise mutualization:} A shared platform enables optimization of a small set of resources, which may be easier to manage than more decentralized patterns.
\end{itemize}

We now explore how \sys can be defined.

\subsection{Time-Sharing units}

We consider computing resources (\gls{cpu}, memory, accelerators...) as the fundamental units that should be considered for time-sharing.

Taking the server, as a coarser-grained unit, as the perimeter for analysis may offer a more optimistic view. 
Indeed, many servers are shared, either sequentially in \gls{hpc} \glspl{dc} or concurrently through spatial sharing in cloud environments, as previously discussed.

However, this perspective introduces a bias, as it overlooks the continual increase in server capacity: the number of transistors per microprocessor has not declined in over 50 years~\cite{rupp2021microprocessor}.

From our standpoint, a cluster of servers cannot be considered smaller simply because the number of servers has decreased.
Instead, attention should be given to the overall computing capacity, which may have increased if the cluster uses more powerful components.
Focusing on computing resources rather than servers avoids naive strategies that suggest renewing hardware solely to benefit from growing computational capabilities.

We argue that attention should be given to the \textbf{envelope} allocated to tenants to support their applications.
Envelopes serve as the interface between software requirements and hardware capacity, and thus, they link application functionalities to their physical impact.
Reducing the per-envelope resource allocation may be the only sharing principle that meaningfully contributes to lowering the total computing capacity.

When reasoning in terms of baselines, \glspl{vcpu} serve as a useful reference point. 
They represent a virtual abstraction of physical cores, whose characteristics have remained relatively stable due to a plateau in single-thread performance and clock frequency.
Recent increases in transistor counts are now mainly due to the proliferation of physical and logical cores rather than improvements in individual core performance~\cite{rupp2021microprocessor}.

\subsection{Time-Sharing policies}

Time-sharing distributes the usage of a given resource over time across multiple applications—a situation we consider more beneficial than assigning dedicated resources per application.
However, time-sharing can be interpreted in various ways. 
In this subsection, we explore several of these interpretations.

Table~\ref{tab:classification} presents a classification of different high-level policies.
We distinguish between two levels of services: virtual resources, where the client manages their own software within a virtualized environment; and functional units (a term borrowed from life cycle analysis~\cite{10292179}), where the hosting provider manages the software to fulfill a specific service need.
In both cases, time-sharing can be sequential or concurrent, leading to different inconveniences.

In the case of sequential time-sharing at the virtual resource granularity, the envelope is not immediately started or scaled, but instead queued until sufficient resources become available. 
This can delay deployment or restrict vertical elasticity. 
Concurrent time-sharing at the virtual resource level allows multiple envelopes to run simultaneously on shared physical resources, with the hypervisor’s scheduler multiplexing execution. 
While this improves utilization, it may also introduce performance contention.

When time-sharing is applied at the functional unit granularity, the provider directly manages software components or services. 
Sequential time-sharing means that requests or tasks are processed one at a time using a limited resource pool, potentially introducing execution delays. 
Concurrent time-sharing at this level involves serving multiple tasks in parallel within the same application environment—common in multi-user SaaS platforms—but this can also degrade performance due to contention over internal shared resources, including hardware components (e.g., CPU, memory) and shared software services (e.g., application threads, databases, or middleware).

Note that this classification focuses more on the hosting point of view (macro perspective).
Unlike micro-level strategies, which require the coordination and participation of all actors, macro-level approaches can be managed by a single entity that centralizes the necessary expertise.

\subsection{Time-Sharing resources}

\begin{table*}[hbt!]
\caption{Examples of Time-Shared Policy Applications for Different Services}\label{tab:examples}
\centering
\begin{tabular}{|p{5cm}|p{11cm}|}
    \hline
    \bf Service Description & \bf Applicable Sharing Paradigm(s) \\
    \hline
    Personal website (e.g., portfolio, blog) & Sequential time-sharing at the functional unit level; low-priority queue with long cache lifetimes. \\
    \hline
    Small webshop (few daily transactions) & Concurrent time-sharing at the virtual resource or functional unit level. \\
    \hline
    Large-scale webshop & Requires more static spatial allocation; hybrid time-sharing with partial resource dedication. \\
    \hline
    Drive service (e.g., file syncing) & Concurrent time-sharing at the functional unit level with controlled replication and compression. \\
    \hline
    Governmental e-service portal & Separation of static and dynamic content: static content served via shared virtual resources; dynamic content using hybrid strategies. \\
    \hline
    Streaming service (audio/video) & Typically requires dedicated bandwidth and latency guarantees; less critical resources (e.g., disk) may be shared; background tasks (e.g., encoding) can be time-shared. \\
    \hline
\end{tabular}
\end{table*}

All resource usage can be optimized.
An algorithm can improve its CPU efficiency by reducing complexity, limit memory consumption through better data structures, or eliminate bottlenecks to accelerate the use of GPU resources.
Trade-offs often emerge in the process—an increase in the use of one resource may lead to a reduction in another (e.g., memory vs. CPU in cache-based architectures).
However, this paper focuses on what comes after optimization.

After optimizations, the next step is sharing—pushing resource utilization below the granularity achieved by a single optimized application (e.g., CPU time slices, memory pages, CUDA cores).
As previously discussed, this sharing can occur either sequentially (by accepting execution delays) or concurrently (by accepting performance degradation).

The CPU core is the most straightforward resource to share, as the kernel already distributes its time efficiently between applications (using, for example, \gls{cfs}~\cite{10.1145/1400097.1400102} or \gls{eevdf}~\cite{stoica1995earliest} on Linux).
Improving sharing at this level is more of an orchestration problem: how many more jobs can or should be deployed?

Memory sharing is more delicate.
Software reserves memory pages, and the total reserved by software corresponds to its \gls{rss}.
However, not all reserved pages are actively used; the actively accessed subset is referred to as the \gls{wss}.
While \gls{rss} may allow for page deduplication and sharing (e.g., through kernel same-page merging~\cite{ksm}), sharing the \gls{wss} is more complex.
Memory constraints can be introduced via sequential sharing strategies (e.g., hot-plug~\cite{hotplug}), or through techniques such as deallocation (e.g., ballooning~\cite{180153}), memory compression (with associated performance degradation~\cite{zram}), or swapping (also incurring performance penalties).

\glspl{gpu} can be shared relatively easily through time-slicing of their compute capacity.
Recent developments in GPU partitioning (e.g., NVIDIA vGPU, NVIDIA MIG) and cooperative applications (NVIDIA MPS) support concurrent sharing.

Network resources can be shared at both hardware and protocol levels.
Technologies such as SR-IOV (Single Root I/O Virtualization) and software-defined networking (SDN) allow for bandwidth partitioning, traffic shaping, and QoS enforcement between tenants.
Sequential sharing can also be considered, for example, by scheduling large data transfers during off-peak hours.
However, maintaining low latency and throughput guarantees across tenants remains a challenging orchestration problem.

Disk storage is less amenable to time-sharing since its primary role is persistent data retention.
The discussion is often more relevant at the infrastructure level, where the trade-offs between SSDs, HDDs, and tape storage are involved.
These technologies differ not only in read/write performance but also in their environmental impact~\cite{simon2024boaviztapi}, underlying that different access frequencies should lead to different solutions.

FPGAs are inherently suited to sequential sharing as they can be considered as a physically programmable system.

\subsection{Choosing service envelopes}

The previously described time-shared hosting envelopes, managed by hosting providers, present different trade-offs in terms of responsiveness, resource availability, and operational complexity.
We now provide insights into how these envelopes can be matched with various types of digital services, helping developers assess their suitability in practical scenarios.

Specifically, Table~\ref{tab:examples} outlines examples of services and potential sharing paradigms.
These paradigms refer to how time-sharing is applied (e.g., sequential or concurrent), and at what level (e.g., virtual resource or functional unit), depending on the needs and constraints of each service.

This mapping is indicative rather than prescriptive.
In practice, more ambitious implementations are possible, such as separating static and dynamic components of services to apply distinct sharing strategies, or designing non-holistic envelopes that allow different resources to be shared according to different policies, depending on workload characteristics.

\subsection{On other impacts}

Most demands in \gls{ict} have a tangible impact in terms of resource consumption, even those not directly tied to the hosting envelope.
For instance, the expectation of fast \gls{vm} provisioning requires maintaining a pool of idle server resources to absorb demand variability~\cite{jacquet2023cloudfactory}.
Another example is elasticity: while scaling an application up or down improves performance responsiveness, it typically requires over-provisioning when reduced delay is expected, which again increases resource consumption even when idle, something that may be overlooked in geo-scheduling strategies~\cite{10.1145/3698038.3698542}.
Finally, redundancy increases the overall hardware footprint and energy usage, and is required when pursuing high availability, robustness or resiliency, even outside the \gls{ict} realm~\cite{hamantAntidoteAuCulte2023}.
All those impacts should not be forgotten while trying to reduce the computing capacity.

\section{Potential implementations}\label{sec:implem}

\begin{figure*}[hbt!]
    \centering
    \includegraphics[width=\linewidth]{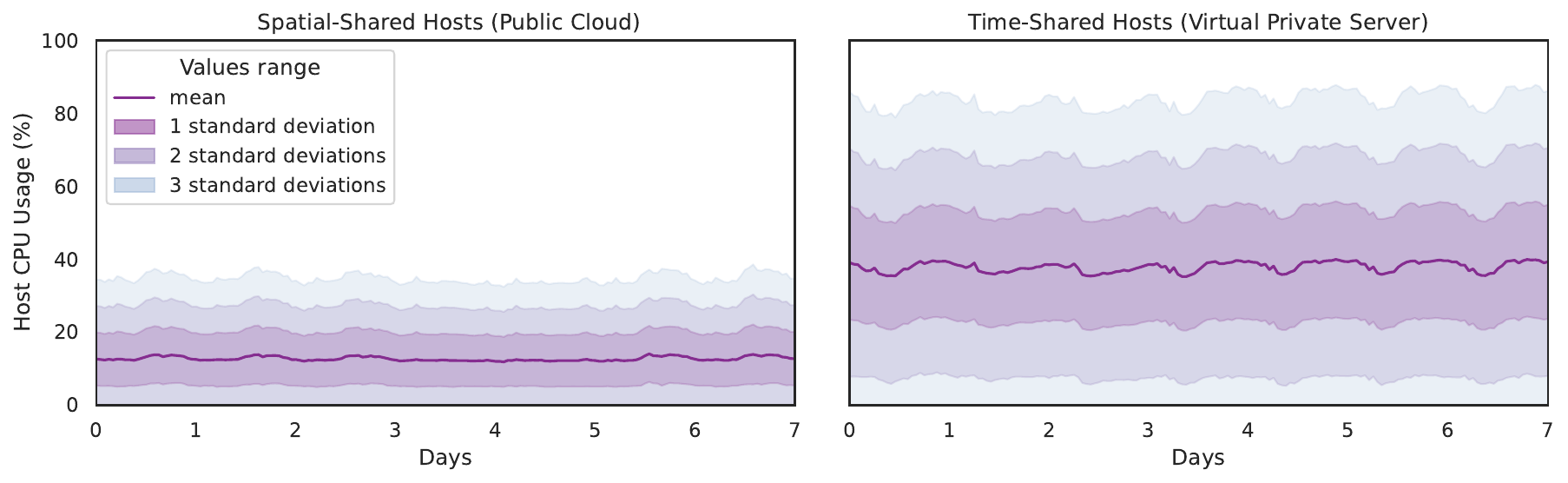}
    \caption{Comparison between different cloud products' usage of host CPU resources in OVHcloud context through a typical week}
    \Description{Comparison between different cloud products' usage of host CPU resources in OVHcloud context through a typical week}
    \label{fig:persp:cloud}
\end{figure*}

In this section, we explore how \sys could be implemented by hosting providers.
In practice, digital services are operated by a wide variety of actors, from large cloud hyperscalers to small web-hosting associations.
To reflect this diversity, we consider both industrial-scale and community-driven infrastructures, highlighting the range of possible implementations and trade-offs.

\subsection{A cloud provider perspective}

We partnered with OVHcloud, one of the largest European cloud providers, operating more than 43 data centers worldwide, to explore time-sharing opportunities in their context.

Our study focuses on first analyzing the impact of resource sharing across different cloud products, selecting two offerings that implement distinct sharing paradigms.

We then explored how the utilization of physical resources could be enhanced within the framework of existing cloud models.

\subsubsection{Current situation}

As previously mentioned, OVHcloud offers a variety of cloud products, ranging from cost-effective instances—referred to as \glspl{vps} in their portfolio—to more premium offerings, known as \gls{pci}. We propose analyzing the resource usage patterns of both product lines.
We use the \gls{cpu} as a proxy for cloud server activity.
Among the various resources (memory, disk, network), we consider the \gls{cpu} to be the most representative, as most types of workloads ultimately rely on \gls{cpu} usage.

Figure~\ref{fig:persp:cloud} presents a unique overview of two distinct clusters: one composed of \glspl{vps}, the other of \glspl{pci}, each consisting of over 1,000 servers.
We report the \gls{cpu} usage of hosts over a week in March 2025 using the mean, along with one, two, and three standard deviations from the mean, capturing approximately 68\%, 95\%, and 99.7\% of the values, respectively, reflecting the Gaussian-like distribution.

A diurnal pattern is evident, particularly within the standard deviation bands, showing that resource usage rises during the day and decreases at night. 
However, the overall variation is relatively modest, suggesting that much of the computing capacity is not utilized for human interactions (e.g., websites). 
As a result, reduced performance would likely have a low impact on user experience. 
Moreover, a significant portion of the resources remains underutilized throughout the week in both scenarios.

The two clusters exhibit distinct behaviors, which reflect their underlying resource-sharing paradigms.
Typically, the cluster optimized for time-sharing allows for increased concurrency of jobs, leading to improved resource usage, with a clear distinction when compared to the \gls{pci} usage patterns.
To the best of our knowledge, this paper is the first to illustrate the scale of resource usage improvements enabled by time-sharing in an \gls{iaas} context, made possible by OVHcloud’s dual product lines. 
When reasoning around the mean, \glspl{vps} enables the use of $3.0$ times more physical computing resources per host, drastically reducing the per-instance physical allocation.
Note that this ratio differs from the applied oversubscription ratio, due to factors such as non-allocated resources, hardware performance, scheduling policies, \gls{vm} size distribution, background services running on the host, and other operational considerations (e.g., maintenance activities).

\subsubsection{Leads}

From our observations on CPU usage, time-sharing appears to be a promising strategy to pool the unused margin between \glspl{vm}—that is, the gap between the resources provisioned and those used at a given time.
However, current solutions can be further improved.
We identify several avenues to enhance resource sharing in cloud environments, building on the taxonomy presented earlier. 
The scope of these actions depends on the product range, which entails varying degrees of provider involvement.
We therefore explore opportunities ranging from virtualized resources (in an \gls{iaas}-like context, where clients manage the software stack of their \glspl{vm}) to higher-level functional units (where most of the resource management is delegated to the cloud provider).\\

In the context of virtualized resources, \textbf{sequential time-sharing} policies have notably been investigated through the concept of Harvesting \glspl{vm}~\cite{10.1145/3447786.3456225}. 
These \glspl{vm} start with minimal guaranteed resources (e.g., a single core) and can opportunistically acquire more if available on the host (in terms of both \gls{cpu} and memory). 
While this approach remains opportunistic, offering resources only when idle, it introduces a hybrid model: some resources are guaranteed, while others are elastic and shared. 
This sequential sharing model can thus serve to absorb demand peaks. 
It also links to the idea of dynamic elasticity—whether through vertical or horizontal scaling—but with a twist: in a constrained setting, resources might not be available when requested, a paradigm less explored.

\textbf{Concurrent time-sharing} policies of virtual resources have been approached from multiple angles. 
Several works focus on estimating the optimal oversubscription level—i.e., how many \glspl{vcpu} can reasonably share a single physical core—using runtime monitoring~\cite{10.1145/3132747.3132772,10.1145/3447786.3456259,DBLP:journals/tsusc/JacquetLR24}. Hybrid models have also been proposed, where only a subset of a \gls{vm}'s \glspl{vcpu} participate in concurrent sharing~\cite{DBLP:conf/ccgrid/JacquetLR24}. 
The benefit of co-hosting \glspl{vm} with different levels of resource sharing has also been highlighted as a means to reduce resource fragmentation~\cite{DBLP:conf/cluster/JacquetLR24}.

Virtual resources are managed in a black-box perspective by the provider.
In such a setting, performance degradation caused by advanced sharing techniques has been studied, both in terms of identifying optimal sharing levels and developing hybrid strategies. 
We believe these hybrid approaches can be extended further for virtual resources:

\begin{itemize} 
    \item \textbf{Enhanced Inference:} Improved inference of internal workloads—either through more intelligent observation or increased communication between the hypervisor and the \gls{vm}—could enable dynamic adjustments of the sharing level as workloads evolve. 
    \item \textbf{Modular Resource Management:} Beyond CPU and memory, other resources could benefit from time-sliced management. Identifying how best to expose such modular, time-shared resources in an \gls{iaas} setting remains an open question. Current credit-based market models (where \glspl{vm} earn credits by not using resources and spend them during bursts~\cite{burstazure,burstaws}) could be revisited in this light.
    \item \textbf{Improved security:} Although side-channel attacks are technically complex, they remain a relevant concern in multi-tenant environments with increasing resource sharing. Potential mitigation strategies include aware scheduling, attack detection mechanisms, use of physical enclaves, and other methods.
    \item \textbf{Footprint Feedback to Users:} The success of any advanced sharing model relies not only on system efficiency but also on user perception. Communicating resource savings and their environmental benefits could incentivize users to adopt more frugal, sustainable configurations. 
\end{itemize}

Note that all these leads can initially be applied to non-critical systems (such as microservices that are not a bottleneck), where the impact of resource sharing is minimal. 
In contrast, applying the same strategies to critical or complete systems introduces trade-offs, with the impact depending on how much sharing is enforced and how performance degrades.\\

Functional units, however, present a different perspective. 
In these models, the client consumes a service, delegating most of the technical management to the provider. 
A prominent example is \gls{dbaas}, where users access database functionalities while the underlying \gls{dbms}, storage, and compute resources are fully managed by the cloud operator. 
This abstraction enables a more accurate understanding of the functional need, opening the door to more fine-grained and efficient optimizations.

A \textbf{sequential time-sharing} policy with this service aligns with the \gls{faas} paradigm, where functions are queued and executed in response to specific triggers, drawing from a shared resource pool. 
While cloud providers typically seek to minimize start-up latency, it is conceivable to intentionally limit the resource pool, thus accepting longer wait times based on priorities, market-based rules, or other fairness strategies.

\textbf{Concurrent time-sharing} policies of functional units would involve serving multiple tenants from the same software stack (e.g., multi-tenant databases), potentially in oversubscribed settings. 
These optimizations are inherently product-specific but aim at the same goal: minimizing physical resource usage by pooling static overhead (e.g., base memory footprint, redundant system processes) across clients. 
As with virtualized resources, a progression can be envisioned: starting with non-critical workloads, then sharing auxiliary services, and ultimately applying slowdowns to full systems if acceptable.

Once again, we believe hybrid strategies—combining opportunistic elasticity with guaranteed service levels—hold the most promise. 
Several directions emerge for functional units:

\begin{itemize} 

    \item \textbf{Remove functionalities:} Managing functionalities at the functional unit granularity gives the cloud provider greater visibility and control over what can be avoided, reduced (or delayed), and offsettable impacts. Certain background processes, redundancy mechanisms, or optional features (e.g., real-time replication, aggressive caching, or analytics modules) could be selectively deactivated for low-priority services or clients willing to accept a leaner setup, freeing up resources while preserving core functionality.

    \item \textbf{Tame performance loss:} A key success factor is the ability to contain the performance degradation in a predictable and controlled way. Equally important is communicating the environmental benefits of frugal configurations to clients, encouraging adoption.

    \item \textbf{Communicate gains:} Just as important as the savings themselves is making them visible to users. Showing clear metrics (e.g., energy saved, carbon footprint reduced) can transform frugality from a technical trade-off into a value proposition for sustainability. 
\end{itemize}

\subsection{A collective hosting provider perspective}

Let us turn our attention to community-driven digital hosting solutions outside of the cloud paradigm.
We will specifically take interest in Deuxfleurs~\cite{deuxfleursDeuxfleursFabriquonsInternet2025}: a French association that provides digital services (e-mail, static websites, videoconferencing, collaborative editing, etc.) to the civil society, including individuals, collectives, and enterprises. 
This endeavor—to offer end-user digital services as a non-commercial commons—is gaining momentum. 
In France, the CHATONS~\cite{chatonsCollectifHebergeursAlternatifs2025} federation now brings together around 90 such hosting collectives.
In this landscape, Deuxfleurs' originality lies in its infrastructural design choices: a production cluster of only 7 second-hand desktop computers---distributed among the households of 3 of their members using domestic fiber optics connections---supplies the totality of their services.

Deuxfleurs developed a distributed object store tailored for low-end workers communicating through high-latency WAN links, coined Garage~\cite{thegarageteamOpensourceDistributedObject2025}.
It tolerates the disconnection of a full availability zone (out of three) without interrupting its operation.
Garage notably backs the \textit{static website hosting service}, making it as resilient and available as cloud offerings.
We focus our analysis on this service.

%

\subsubsection{Current situation}
The \textit{material envelope} of Deuxfleurs' infrastructure is salient: the 7 desktop computers constituting the production cluster.
The association intends to keep at this envelope for the foreseeable future, and to organize the \textit{sharing} of these material resources by constraining and optimizing usages, instead of elastically scaling its envelope.
Deuxfleurs currently hosts more than 500 static websites.
It estimates that its network and storage budget are only used at 4\% and 6\% of their capacity, respectively. 
Because only static websites are served, the CPU footprint (mostly caused by HTTPS cryptography) is negligible.

At the time of writing, a simple quota mechanism ensures a fair sharing of the storage space among users: each website is allowed 50\;MB of storage, which can be increased up to 200\;MB in autonomy by its owner. 
The motivation for such a mechanism is to raise awareness about the material impacts of our digital actions.
If a use-case requires more storage (e.g., a graphical artist or musician's portfolio), a custom quota for the website is proposed after a collective discussion.

\subsubsection{Leads}

We now outline key challenges associated with the mutualization of computing resources in collective hosting and how Deuxfleurs envisions addressing them.

\paragraph{CPU time-sharing}
Despite their unmatched simplicity, static websites lack the interactivity that once popularized Web 2.0, such as participatory features (forms, comments, likes), access control (e.g., for administrative back offices), advanced search capabilities, and more.
All of these require server-side code execution, i.e., CPU resources, that Deuxfleurs would like to propose in its future service offerings, at the granularity of the \textit{function}.
To ensure fair sharing of compute resources among different tenants, Deuxfleurs is considering implementing a function execution queue based on sequential time-sharing, where each function’s priority is weighted by the requesting website’s \textit{karma}: the higher the karma, the higher the execution priority.
Karma would decrease upon each execution and gradually replenish over time.
When a website's karma would reach zero, its requested executions would have the lowest priority, and could even be dropped in case of overflow. 

\paragraph{Incentivize responsible behavior on shared resources}
The association recently witnessed slow time-to-first-byte on website GET requests.
This issue stemmed from two main causes: a failing disk limiting the throughput of the object store (prompting an urgent replacement), aggravated by bursts of excessively high rates of PUT requests (i.e. websites updates).
These bursts are caused by the careless implementation of some automation deployment tooling, typically re-uploading a whole website instead of cherry-picking what needs synchronization. 

A feedback control algorithm (akin to TCP's AIMD~\cite{chiuAnalysisIncreaseDecrease1989}) is being developed to limit the consequences of such administrative traffic on the websites' responsiveness. 
In this time-shared policy, the acceptance window for PUT requests will be voluntarily shrunk under heavy load and gradually expanded as congestion decreases..
This approach will naturally penalize poorly designed automation pipelines that flood the system, encouraging administrators to adopt more efficient, minimalist deployment strategies.
Which is precisely the goal: a limited, shared infrastructure must be operated with respect and care.

\paragraph{Monetary Cost-Sharing}

Computing resources are inherently tied to cost-sharing mechanisms. 
While the business model of cloud computing providers is clearly established and commercially optimized, operating similar infrastructures as a non-profit organization presents unique challenges. 
Associations must explore and adopt alternative, community-oriented economic models to ensure sustainability.

Deuxfleurs invested 786\;€ for Garage in 2024~\cite{deuxfleursFinancialReport20252025}. 
Hard drives \& SSDs represent the majority of this expenditure: they are the most consumable hardware parts of the infrastructure, and must regularly be replaced.
The electrical cost of the servers is unmonitored and graciously paid for by the hosting volunteers, but it is estimated to represent several hundred euros annually.
Users' contributions \& donations currently cover all of the association's recurring costs.
However, in 2024, Deuxfleurs began offering hosting services to web agencies, who willingly pay a reasonable fee for the association's resilient and responsible static website hosting.

The ambition is to keep Deuxfleurs a community-driven endeavor, with infrastructural decisions being made by its members \& users on an equal footing.
To advance as a serious digital governance proposition, the association fosters economic activity developing on top of its commons, and intends to employ its financial resources to support future software development, public advocacy, design as well as design and arts initiatives.
%




\section{Conclusion}\label{sec:conclusion}

In this paper, we advocated for a more efficient sharing of computing resources within hosting infrastructures. 
While spatial sharing—allocating distinct resource pools to different clients—is a common practice, temporal sharing remains largely under-exploited.

Yet, time-sharing offers a compelling paradigm: allocating a resource over time to multiple workloads is often more efficient and sustainable than dedicating separate resources to each use case.
Although time-sharing was a foundational principle in the early days of \gls{ict}, its prominence has faded with the proliferation of abundant computing capacity.

We argued that the future of \gls{ict} is likely to be more constrained—due to environmental, economic, and physical limits—and that a revival of time-sharing strategies will become increasingly relevant. 
We have explored how modern time-sharing can be reimagined and applied within contemporary hosting computing models, ranging from infrastructure-level virtualization to higher-level functional services.

Finally, we outlined concrete directions for implementing time-sharing in both industrial cloud platforms and community-driven hosting initiatives, as a means to foster more frugal, efficient, and sustainable digital infrastructures.

\begin{acks}\label{sec:ack}
This work was partially supported by Mitacs and OVHcloud under project IT42864.
\end{acks}

\bibliographystyle{acm}
\bibliography{sharing} 

\end{document}